# DETECTING ERRORS IN SPREADSHEETS

Yirsaw Ayalew, Markus Clermont, Roland T. Mittermeir
Institut für Informatik-Systeme, Universität Klagenfurt
Klagenfurt, Austria
Email: {yirsaw, mark, mittermeir}@ifi.uni-klu.ac.at

**ABSTRACT**

*The paper presents two complementary strategies for identifying errors in spreadsheet programs. The strategies presented are grounded on the assumption that spreadsheets are software, albeit of a different nature than conventional procedural software. Correspondingly, strategies for identifying errors have to take into account the inherent properties of spreadsheets as much as they have to recognize that the conceptual models of "spreadsheet programmers" differ from the conceptual models of conventional programmers. Nevertheless, nobody can and will write a spreadsheet, without having such a conceptual model in mind, be it of numeric nature or be it of geometrical nature focused on some layout.*

## 1 INTRODUCTION

Spreadsheet systems are the most widely used and the most popular end user systems. Hence, spreadsheets (we might refer to them as "spreadsheet programs") are an important basis for far reaching decisions in almost any field of a modern society. Studies on the quality of spreadsheets resp. spreadsheet based decisions show, however, that there is a substantial divergence between significance and care in this area [3, 6, 12, 13, 14, 15, 16].

Nardi and Miller [10, 11] discussed the characteristics spreadsheet languages provide for end-user programming. Among them is the property that spreadsheets shield users from low-level details of traditional programming. They allow users to think in terms of tabular layouts of adequately arranged and textually designated numbers. Thus, they appear to users as analogous to pencil and paper. The computing model upon which they finally rest is so hidden from the users that the term "programming" seems inappropriate and the term "testing" simply inapplicable.

As professionals we have to recognize, though, that below the surface, spreadsheets are programs. They are even special programs from the perspective that the placement of code is dependent on the layout of the result. – A fact that seems to reduce complexity at first sight (and it does so in simple cases), but that might become a burden in complex situations and specifically during modification. Thus, given the factual importance of spreadsheets due to the importance of the decisions based upon spreadsheet computations, very conventional considerations for software quality need to be considered. These considerations might encompass testing as much as they might encompass design or maintenance and configuration management. However, in using these technical terms, we must not forget that the spreadsheet user does not consider him-/herself as a programmer. (S)he is an end-user who does not want to be bothered with technicalities of the world of programming. In order to become successful, approaches to improve quality control for spreadsheets have to avoid conventional programming- or software engineering jargon. They rather have to link directly to the conceptual structures, spreadsheet users have readily available.

In this paper we will therefore first try to highlight some commonalities and some differences between spreadsheets and conventional algorithmic software. We then give some definitions of the basic terms needed for a focussed discussion about errors in spreadsheets. In section 4, a framework for the classification of spreadsheet faults is described on the basis of some prototypical errors. Finally, in section 5, two complementary approaches to alleviate quality problems in spreadsheet programs are outlined.

## 2   SPREADSHEETS AND SOFTWARE: WHAT'S DIFFERENT?

*Software is written in a professional manner by Professionals; Spreadsheets are written by End-Users!* While this statement is true on some face value, it raises wrong connotations. Software professionals, if working professionally, will build their products based on design that is based on some conceptual model or specification linking the application problem to an algorithmic solution with the algorithm usually considering also certain computer idiosyncracies (input/output being not the least among them). Spreadsheet-writers are end-users. As such, they are not programming professionals. However, they are professionals too, professionals in their application domain. In this capacity, they – like anybody else who writes something meaningful – do express themselves based on some conceptual model whenever they express their problems/solutions in writing. Of course, this applies also when they express themselves in writing a spreadsheet. The only difference to the software professional is that their model is not related to programming concepts. It relates application aspects to two-dimensional (tabular) arrangements of numbers interspersed with explanatory text. The numbers are further conceptually interrelated by either one of the following situations:

- The given number is the result of a computation of some other numbers placed (or to be written later) at a given location.

- The given number is part of a set of numbers playing conceptually the same role. This "same role" is generally expressed by geometrical proximity (physical area). However, we will later identify cases where this conceptual embracement cannot be expressed by geometrical proximity (logical area).

Spreadsheet experts will recognize that the two cases mentioned are not comprehensive. However, we claim that they cover most of the territory, at least most of the territory "non-expert" end-users are familiar with.

As spreadsheet systems are easy to use, do not require much training in formal methods of designing and programming, and show – in contrast to conventional programs – the results of the effort while the development effort is still in progress, they are also written in a style different from conventional software. There is a notion of immediate feedback [5] once the contents of a cell is specified. This easy way to quick feedback leads to a development style of trial & error, cutting & pasting, copying & modifying; a mixture that has to be horrifying for an orderly software methodologist.

Figure 1 shows these aspects. Given these considerations, it becomes obvious that irrespective of the true nature of spreadsheet-"software", conventional wisdom on software testing (c.f. [2, 9, 17]) does either not apply at all or applies only to a limited extent. It applies specifically from the perspective though that spreadsheet computations are basically numerical computations. We will come back to this property in section 5.2.

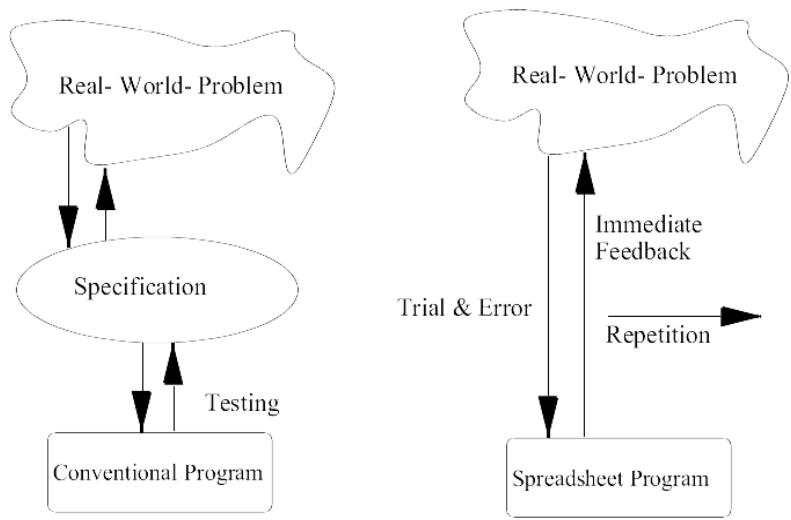

**Figure 1**: Conventional program vs. spreadsheet program development process

Hence, rather than banking too much on preaching the gospel of "designing first" and establishing a quality improvement cure on observing this commandment, we base our approach on the very nature of existing spreadsheets and existing processes of how spreadsheets are written. In this paper, we focus specifically on "spreadsheets as they are". This leads us to discuss in the sequel model visualization and plausibility testing. With model visualization, we bank on end-users transforming their problems/solutions into two-dimensional structures. Visualization will highlight irregularities in this transformation. For plausibility testing we rely on the end-users gut feeling for meaningful boundaries of the data (numbers!) treated in a spreadsheet. Discussing strategies for ensuring spreadsheet quality dynamically (focusing on spreadsheet evolution) would be beyond the scope of this paper.

Before delving into both of these areas, we will proceed by defining some key terms needed in the further discussion and mentioning some prototypical faults in spreadsheets and their related categories.

## 3 SOME TERMINOLOGY

Since the term "spreadsheet" itself is overloaded, we explain below the semantics attached to spreadsheet related terms in this paper.

A *cell* is the atomic unit of a spreadsheet and can have five states: (a) it can be *empty*, (b) it can hold a *constant value* that is supplied by the programmer of the spreadsheet, (c) it can hold an *input value* that is supplied by the user of the spreadsheet, (d) it can hold a value that is *calculated* by a formula, or (e) it can hold a *label* that describes the contents of a set of other cells.

A *formula* is a mathematical expression, containing cell references, operators, functions[1], and constant values. At least one cell-reference is expected to be included in the computational expression of the formula. A formula yields exactly one result and is free of side-effects.

---

[1] A function is a built-in formula supplied by the spreadsheet system

A *cell reference* is a reference to another cell's value which is either *relative* or *absolute*. The address of the referenced cell is given with a pair of coordinates, in the first case the origin is the referencing cell, in the latter the upper left corner of the spreadsheet.

An *area* is a set of related cells. If the cells are spatially neighbors and the area is marked by the programmer, we refer to it as physical area. A physical area usually serves as the input for a grouping function, like *SUM, MAX or AVG*. If the relation originates from similarities of the data-manipulation or from the way of creation (i.e *copy and paste*), we use the term logical area. We require the cells in a physical area to be also spatially adjacent, for cells in a logical area, no such criterion is defined. The logical area is used to describe a kind of conceptual cohesion between cells. If we cannot figure out the way the cells were created (e.g. *copy and paste* of same source), we have to employ certain heuristics that are based on the similarity of the references and formulas to group cells into logical areas.

A *spreadsheet* is an *n*-dimensional matrix of cells. Each cell is uniquely identified by *n*-coordinates. If *n*=2, as in the standard case, a cell is uniquely identified by its row and column address.

A *Spreadsheet Core Language (SCL)* is a set of language constructs to describe the data-flow (cell references) and the data-manipulation (formulas) in the spreadsheet program. Functional properties of the spreadsheet are expressed by SCL. The *copy and paste* primitives are also considered to be part of the SCL, if they are used in a context with logical areas.

A *Spreadsheet Language (SL)* is the SCL together with constructs for manipulation of the layout of the spreadsheet.

A *Spreadsheet Program (SP)* is the specification of data-flow between cells, data-manipulation in cells, and of the values of constant cells.

A *Spreadsheet Instance (SI)* is a spreadsheet program, where all input cells have certain values. A spreadsheet program can be instantiated multiple times. By changing one of the input values, the spreadsheet instance of a certain spreadsheet program is transformed into another spreadsheet instance of the same program.

A *Spreadsheet System* is an integrated environment, where spreadsheet programs can be created, instantiated and edited. The spreadsheet system interprets a specific spreadsheet language.

## 4 FAULTY SPREADSHEETS AND ERROR CATEGORIES

In this section we describe a framework that enables us to categorizes errors by their association to spreadsheet concepts. We define three categories of errors that are associated with physical areas, logical areas, and general errors. Each of the categories will be briefly described and the reasons for the assignment of a certain error to its category will be given. The examples shown try to demonstrate how an error originates. Of course, all the problems shown could have been solved in another way, without an error occurring.

A classification scheme should address the types of most frequent important errors. In addition, the effectiveness of error prevention and detection techniques can be evaluated, pro-vided that there is a taxonomy of errors which indicates the types, frequency, and possible causes. However, as Beizer [2] indicated, there is no universally correct way to categorize faults. A given fault can be put into different categories depending on the view of the tester and the source of the error. For example, typing "+" indstead of "-" in a given formula might be a typographical error or result form misunderstanding the necessary arithmetic.

Some classification schemes are available for spreadsheet errors. Panko and Halverson [15] offer a taxonomy that consists of three major categories of errors: mechanical, logic, and omission errors. Mechanical errors refer to typographical and positioning errors. Logic errors are misunderstandings of the logic of the necessary algorithm to be used in a formula. Omission errors are a result of leaving out something needed in the program. This classification is mainly based on the causes of the errors. A more general classification scheme containing Panko and Halverson's scheme is given by Rajalingham et al. [20].

In their experimental study, Saariluoma et al. [22] categorized spreadsheet errors in two basic types: location and formula errors. Location errors are what is commonly termed as misreference errors. Saariluoma et al. indicated that these errors are typical in spreadsheet programs. Formula errors contain typographical errors in formula components and what they call mathematical mistakes. Mathematical errors are a result of the inability to define the necessary mathematical expression in a formula. The main errors in this scheme are typo, misreference, and mathematical errors.

Unlike other classification schemes, we do not want to categorize the errors by their cause, but rather by the spreadsheet concept they seem to be associated with. In our further considerations we do not make a difference between logical, mathematical, or typographic errors, because from the error itself we cannot resolve its cause.

### 4.1 Category 1: Physical Area Related Errors

Errors that are typical to physical areas normally deal with missing values in the area or values of the wrong type somewhere in the area. We call this error *reference to a blank cell* resp. *reference to a cell with a value of wrong type*. In some cases such values are entered on purpose, to achieve a better structure and/or readability of the spreadsheet program. In other cases, these values result from errors.

|    | A        | B            |
|----|----------|--------------|
| 1  |          |              |
| 2  |          | 1. Quarter   |
| 3  |          |              |
| 4  | January  | 140          |
| 5  | February | 200          |
| 6  | March    | 170          |
| 7  |          | 2. Quarter   |
| 8  | April    | 180          |
| 9  | May      | 230          |
| 10 | June     | 100          |
| 11 |          |              |
| 12 | 1. Sum   | =SUMME(B2:B10) |

**Figure 2**: Reference to a blank/wrong typed cell

**Example 1: Reference to a blank/wrong typed cell**

*In Figure 2 the range for the sum spans from label **1. Quarter** down to the last cell of the list. The two label cells are not considered in the sum yet, but there is no hint for the user/programmer that they might influence the sum, if they are changed to a number (e.g to **1** instead of **1. Quarter**).*

Another typical problem of the physical area is the impact on the results if new values are added to the area. If a new value is inserted somewhere in the middle of the physical area, it automatically expands, such that the new value and all old values are still within the area. If the new values are added by appending them to the area, the area does not expand. This leads to the error type of *incorrect range specification*.

Generally, the incorrect range specification problem exists if there are cells outside the physical area that should be part of it. For the user it is not clear that those cells are not part of the physical area any more and it is common for him/her to assume that those cells influence the result of the function applied to the physical area, too.

**Figure 3**: Formation of an incorrect range specification error

**Example 2: Physical Area Specification Error**

*In Figure 3 the user defines a sum over an area of cells. During the lifespan of the spreadsheet program it turns out that more cells are needed for specifying the revenues of the salesmen. This is not a problem for extending Miller's range. But the row appended for Smith is not part of the physical area anymore. The sum-cell C8 does not yield the correct result. However, the reason why the final spreadsheet instance is wrong is not obvious for the user.*

A third class of typical errors is the *accidental deletion of a cell* within a physical area. This leads to the already identified *reference to a blank cell* error. In addition, adding something that should not be present, will have similar consequences.

A fourth class of errors is the *physical area mix up* error. While the previous error categories are grounded on the fact that users hardly distinguish between spreadsheet programs and spreadsheet instances (input has not the distinct role as in conventional programming), this error class is due to the spreadsheet program's property of being a mixture of a problem solving tool and a presentation tool. The problem arises, when two separate physical areas get mixed up. In this case one of them cannot be defined as a physical area by the user anymore. The grouping functions have to be replaced by expressions (i.e SUM by multiple +). For the user it is not obvious that (s)he can specify two physical areas in two columns (see lefthand-side of Figure 4), but that it is not allowed to merge them in one column resp. that the result of the grouping function applied to one of the physical areas is not correct any more.

**Example 3: Physical area mix up problem**

*As shown in Figure 4, the salesman spreadsheet program has to calculate a final sum over all sales and a subsum for each salesman. If the user wishes to place the final sum, the subsum and the sales in one column (i.e. for layout reasons), the final sum has to be replaced by an expression which adds the subsums. If the subsum moves to another cell or another salesman (with a new subsum) is introduced, the user has to maintain the final sum expression. If (s)he forgets it, the final sum becomes wrong.*

|   | A | B | C | D | E | F | G | H |
|---|---|---|---|---|---|---|---|---|
| 1 | Salesman | Date | Sales |   |   | Salesman | Date | Sales |
| 2 |   |   |   |   |   |   |   |   |
| 3 | Miller | 01.4.2000 | 500 |   |   | Miller | 01.4.2000 | 500 |
| 4 |   | 16.4.2000 | 1000 |   |   |   | 16.4.2000 | 1000 |
| 5 |   | 18.4.2000 | 900 |   |   |   | 18.4.2000 | 900 |
| 6 | Subtotal Miller |   |   | 2400 |   | Subtotal Miller |   | 2400 |
| 7 | Smith | 04.4.2000 | 600 |   |   | Smith | 04.4.2000 | 600 |
| 8 |   | 06.4.2000 | 900 |   |   |   | 06.4.2000 | 900 |
| 9 |   | 16.4.2000 | 1000 |   |   |   | 16.4.2000 | 1000 |
| 10 | Subtotal Smith |   |   | 2500 |   | Subtotal Smith |   | 2500 |
| 11 | Total |   |   | 4900 |   | Total |   | 4900 |

**Figure 4**: Physical area mix up problem

### 4.2 Category 2: Logical Area Related Errors

As defined in section 3, a logical area represents some kind of cohesion between cells. Normally a logical area originates from copying the same source multiple times and the user is not aware of the logical area, which a cell belongs to.

A typical error is *overwriting a formula with a constant value*. This error can have many reasons, like rounding errors or unexpected results of the formula. The user simply overwrites the formula result in the cell with a constant value. Of course, this value remains there, even if the values in the formerly referenced cells change.

Another error that is common to logical areas is *copy misreference*. In this case, a constant value or an absolute reference is specified in a formula, instead of a relative reference. This error is generally not noticed until the cell's formula is copied into another cell. If a constant cell is referenced with a relative reference, a similar problem will occur when the cell's formula is copied.

### 4.3 Category 3: General Errors

General errors are not explicitly associated with a physical or logical area. Few of them are made when entering values into input cells while most of them are made during formula definition. An error associated with input cells is only typographical; e.g., a user might type 75 instead of 750. Incorrect use of formats also affects the way a value is displayed. One might format a value as 0.2 % while the intended meaning could have been 20 %. This can happen to both input cells and formula cells. In addition, if a numeric data is formatted as label data, then it might affect the computed value of a formula.

Another group of general errors is made during formula definition. As stated in section 3 a formula may involve cell references, operators, functions, and constant values. An error can be made in any of these components due to typographical errors or inability to formulate the necessary mathematical expression. These errors include operator errors, boundary errors, parentheses errors, and function errors.

### 5 QUALITY IMPROVEMENT APPROACHES

Here, two concepts to increase the quality of spreadsheet programs are presented. The approaches deal with the different classes of errors discussed in the above categorization. Prevention and detection of the variety of errors described above requires different methodologies that are currently not available for spreadsheet systems.

We first discuss *model visualization*. This gives the spreadsheet programmer resp. the spreadsheet user more insight into the structure of the spreadsheet, which is expected to help shortening the trial and error process of creating the spreadsheet and to understand and debug spreadsheets in use. The other approach deals with *interval testing* spreadsheets. It tries to overcome the difficulties resulting from the lack of specification of spreadsheets by introducing interval arithmetic as basic device.

### 5.1  Model Visualization

The fact that spreadsheet models[2] are "buried in the formulas" [6] obviously makes it very hard to understand and to reconstruct the spreadsheet model.

The buried model has to be reconstructed, to enable the developer or tester to see beyond the formulas to the underlying logic and structure. To achieve this we must consider both the dataflow in the spreadsheet (as suggested by [21]) and the static aspects, such as logical and physical areas. The generation of such a representation of the spreadsheet model should be automatic with little or no intervention of the programmer. Once generated, the spreadsheet model can be used for visualization and for the automatic comparison of spreadsheet programs.

The visualization should support different resolutions, from coarse to fine grained, to give the user resp. programmer the possibility to have a look at the spreadsheet program on the level of physical and logical areas and the dataflow between those areas. In a further step there should be a possibility for the user to zoom into certain areas and to get a more detailed overview on formula or cell-reference resolution.

We plan to realize the graphical visualization of the model in a way that is based on the data-flow graph of the spreadsheet (see [1,8]), but also visualizing logical and physical areas. The user should be enabled to navigate in the visualized model as suggested in Storey et al. [23]. These authors suggest a representation which allows zooming into specific areas of a graph, without loosing the overview about the context, using a *fisheye view* (see [4]).

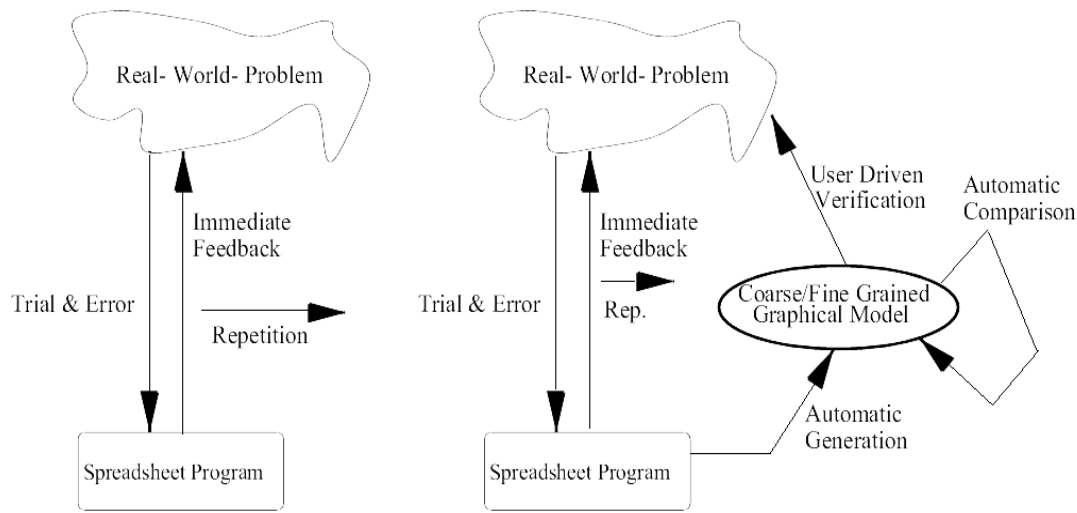

**Figure 5**: Shortening the trial and error process

---

[2]Abstract representation of a spreadsheet program

Our visualized model should serve as a tool for three purposes:

1. Shortening the trial and error process to develop solutions for real-world problems (see Figure 5). We assume that problem understanding is supported by the graphical representation of the spreadsheet model.

2. Understanding of spreadsheet programs that were developed by another programmer.

3. Enabling comparison of spreadsheet programs at the level of the spreadsheet model. This comparison should abstract from values and consider only the model properties, like data-flow, physical and logical areas.

The visualized model will give a representation of physical areas. This gives visual feedback to the user, if there are cells of different types or cells of different conceptual content in the area. A physical or logical area might be visualized as a box, interruptions are indicated using lines of a different color. This visualization should help to control the *reference to a cell with a value of wrong type* problem.

It has also to be checked if there are adjacent cells to the physical area which have the same type as the cells of the area[3]. This might be a hint for the *incorrect physical area specification* problem which can be properly visualized by drawing the required border in a different color.

The *physical area mix up* problem can be resolved by separating the overlapping areas again in the graphical representation. The detection of such overlapping areas, however, is not a trivial problem and further research has to be done on this topic.

By identifying and visualizing logical areas, a concept that is not visually expressed for the user resp. programmer in modern spreadsheet systems, a lot of the problems presented in section 4.2 are already alleviated. Logical areas will often be spatially adjacent, although that is not a necessity. Sporadic interruptions by only a few cells might be a hint for the *overwriting a formula with a constant value* problem. The way of visualization should be similar to visualizing the *reference to a cell with value of wrong type* problem.

## 5.2 Interval Testing

After creating a spreadsheet program for a particular application, it is natural to check its correctness. Spreadsheet programs derve mainly to perform numerical computations. What do people expect to be correct? Usually, one has a gut feeling of the range of reasonable values for each given cell.

Spreadsheet development is based on cells which are to be filled with input values and formulas for computation. For the correctness of a spreadsheet program, every input value as well as every formula should be correct. Actually, many spreadsheet errors are made during formula definition. To judge the validity of the value of a formula cell, we check whether the computation is in the range of expected results. However, the expected behavior of a spreadsheet program is not explicitly specified.

The main duty in testing a program is to detect the existence of a fault in the program. To achieve this one needs systematically designed test cases (using an appropriate test strategy) that reveal faults in the program. By running the program with the test cases and comparing the result with the expected outcome described in the specification or generated by a test oracle[4], the existence of a fault can be detected.

---

[3] If there are cells with values of different types in the area, the correct type can be resolved from the grouping function, which is applied to the area.

Generating a powerful oracle, however, presupposes the existence of a specification [7,18, 19]. Here, we neither have the specification required, nor would spreadsheet developers have the patience and expertise to run a lengthy suite of test cases. Hence, mechanisms need to be devised to approach the power of a test oracle while putting minimal strains on the developers' diligence and insight into complex dependencies. Thus, we must recognize that "testers" of spreadsheet programs are end-users who are not aware of testing theory and hence they are not expected to do testing in the traditional sense. Rather, users of spreadsheet systems are highly dependent on the system's assistance. The fact that control structures are confined to cell contents (and in general used rather rarely if compared to algorithmic programs) allows us to use interval arithmetic as proxy for the services of powerful test oracles.

Figure 6 depicts the test process for a spreadsheet program. Based on the goal of computation and by looking at the input values of cells referenced in a formula, the user, assuming the role of a human oracle, specifies the expected range of computation of a formula in the form of an interval for permissible/expected values.

Each actual value assumed by a cell is a discrete value, either entered by the user as input or obtained as result of a computation by the spreadsheet program. For each of these cells, a range of permissible values has to be given. This is much simpler than generating test cases (which is a very complex process especially for end users) as seen in imperative programs.

The user specifies intervals for those input cells which may assume different values. Those cells which do not assume different values can be represented by an interval of length zero. Therefore, for a formula cell under test, there are two values to be computed and compared (c.f. Figure 6): a value (**d**), computed by the spreadsheet program based on the values of the cells referenced in the formula; a bounding interval (**B**) computed by interval program based on interval arithmetic using the interval values of the cells referenced. The interval program is an equivalent of a spreadsheet program where the values of cells are represented as intervals and the computation is performed based on interval arithmetic. The third value referred to on the right side of Figure 6 is the interval (**E**), the user expects as range for result values.

In order to infer the existence of a fault in a formula cell, the three values **d**, **E**, and **B** which are generated by different sources should be compared. There are two cases to consider.

**case 1: $d \in E$ and $E \subseteq B$**

As the computed interval value of a formula is bounded by minimum and maximum values of the possible computation (by definition of interval arithmetic), the expected interval **E** should lie within the computed interval **B**. Further, the value **d**, computed by the spreadsheet program, should be within the expected magnitude of computation **E**. Hence, in this case, we can say that there is no symptom of fault.

---

[4]A mechanism that predicts the expected behavior of a program based on a specification

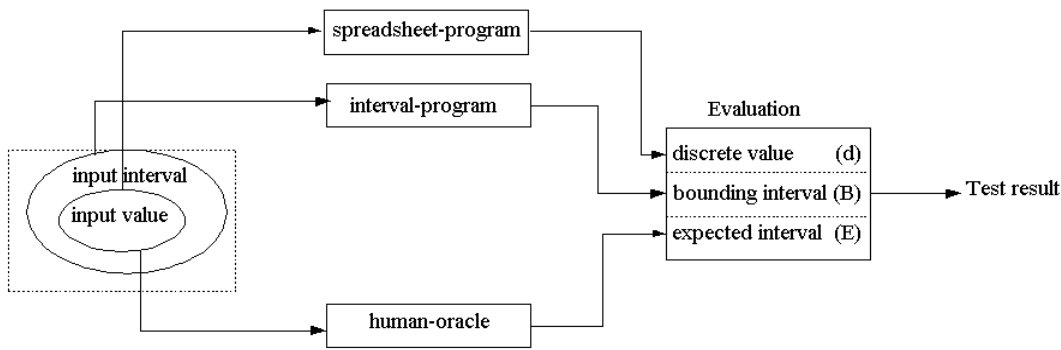

**Figure 6**: Spreadsheet program test process

**case 2: d ∉ E or E ⊄ B**

In this case, there is an indication of symptom of fault. The fault may be in the formula or in the user's perception of expected results. Of course, testing is performed based on the assumption that there is a correct behavior of a program against which the actual result is compared. However, we can not always take for granted that the expected behavior is correct. In the situation where **d ∉ E**, due to some misreferences of cells in the formula or some other errors, the actual result is outside of the range of the expected result. In the second possibility where **E ⊄ B**, faults affect the bounding interval computed for the formula and create a misalignment between **E** and **B**.

This approach is mainly targeted to misreference and incorrect range specification errors. These errors are a result of specifying or selecting a group of cells incorrectly to achieve the desired goal of computation. Generally, we can say that these errors are failures in specifying a plan for a given computational goal. Misreference and incorrect range specification errors are likely to create a misalignment between the values computed by the spreadsheet program, interval program and the expected interval specified by the user. In addition, other errors may also create a discrepancy between the values **d**, **E**, and **B** and could be detected in the process. Once the existence of a fault in a formula is known, the source of the fault may be traced using the data dependency relation between cells established through the formula.

It has to be acknowledged that this form of interval testing plays a dual role. On one hand, it identifies faults in spreadsheet instances, whenever actual values **d** fall outside of the permitted range. On the other hand, the comparison between **E** and **B** is rather a check on the consistency of the user's arithmetic model. Thus, this check can be quite powerful on a much more general level than on the level of a specific spreadsheet instance.

## 6 CONCLUSION

This paper attempts to overcome the tension between the statements *"Spreadsheets are Software too"* and *"spreadsheet-authors are no Programmers"* in order to improve the quality of spreadsheet software.

It is shown that there seems to be no single answer serving as silver bullet. However, a mix of approaches, close enough to the end-users' conceptual model of plausible ranges for values of items as well as visualization of the mapping of conceptual structures to cell arrangements might help to highlight errors of frequently occurring nature.